\documentclass{raa}            
\usepackage[round]{natbib}
\usepackage{graphicx,times}    
\usepackage{amssymb,amsmath}
\usepackage[pass]{geometry}
\usepackage{pdflscape}
\usepackage{multirow}
\usepackage[round]{natbib}

\begin{document}
   \title{A catalogue of H$\alpha$ emission-line point sources in the vicinity fields of M\,31 and M\,33 from the LAMOST survey}
   \volnopage{Vol.0 (200x) No.0, 000--000}      
   \setcounter{page}{1}          

   \author{M. Zhang
      \inst{1, 2, 3}
    \and B.-Q. Chen
      \inst{2} 
      \and  Z.-Y. Huo
       \inst{4}
       \and H.-W. Zhang
      \inst{1, 3}   
      \and M.-S. Xiang
      \inst{4}
      \and H.-B. Yuan
      \inst{5}     
      \and Y. Huang
      \inst{2} 
      \and C. Wang
      \inst{1, 2} 
       \and X.-W. Liu
      \inst{2}
      }

   \institute{Department of Astronomy, Peking University, Beijing 100871, P. R. China; {\it zhanghw@pku.edu.cn}\\
        \and
             South-Western Institute For Astronomy Research, Yunnan University, Kunming 650500, P. R. China; {\it x.liu@ynu.edu.cn, bchen@ynu.edu.cn}\\
        \and
        Kavli Institute for Astronomy and Astrophysics, Peking University, Beijing 100871, P. R. China \\
        \and
             National Astronomical Observatories, Chinese Academy of Sciences, Beijing 100012, P. R. China \\
         \and
            Department of Astronomy, Beijing Normal University, Beijing 100875, P. R. China \\
}

   \date{Received~~2019 xx ; accepted~~2019~~month day}

\abstract{ 
We present a catalogue of 3,305 H$\alpha$ emission-line point sources observed with the
 Large Sky Area Multi-Object Fiber Spectroscopic Telescope (LAMOST) in the vicinity fields of M\,31 and M\,33 
 during September 2011 and January 2016. 
 The catalogue contains 1,487 emission-line stars, 
532 emission-line nebulae including 377 likely planetary nebulae (PNe), 83 H~{\textsc{ii}} regions candidates and 20 possible supernovae remnants (SNRs) 
and 1,286 unknown objects. 
 Among them, 24 PN candidates, 19 H~{\sc ii} region candidates, 10  SNR candidates and 1 symbiotic star candidate are new discoveries. 
 Radial velocities and fluxes estimated from the H$\alpha$ line and those quantities of seven other major emission lines including
 H$\beta$, [O~{\textsc{iii}}]~$\lambda$4959, [O~{\textsc{iii}}]~$\lambda$5007,  [N~{\textsc{ii}}]~$\lambda$6548,  [N~{\textsc{ii}}]~$\lambda$6583, 
[S~{\textsc{ii}}]~$\lambda$6717 and [S~{\textsc{ii}}]~$\lambda$6731 lines
of all the catalogued sources yielded from the LAMOST spectra are also presented in our catalogue. Our catalogue is an ideal starting
point to study the chemistry properties and kinematics of M\,31 and M\,33.
}

   \authorrunning{Zhang et al. }            
   \titlerunning{H$\alpha$ emission-line point sources in M31 \& M33}

   \maketitle

\section{Introduction}           
\label{sect:intro}
The Andromeda Galaxy (M\,31), the closest spiral galaxy and the most dominant member of the Local Group, is always one of the most interesting targets for astronomers.
The Triangulum Galaxy (M\,33), located in the southeast of M\,31 (about 15\degr\ away),  is the third most massive member in the Local Group. 
A large number of work have been carried out to investigate the properties and to conjecture the accretion and evolution history of the stellar substructures and dwarf galaxies in the halo of M\,31 and M\,33 in recent years (e.g., \citealt{Ib2001}; \citealt{Fer2002};  \citealt{Ka2006};  \citealt{Gi2007}; \citealt{Ib2007}; \citealt{Gi2009}; \citealt{Mc2009}; \citealt{Ta2010} and references therein). 
Most of these studies are based on the photometric data.
Due to the large distances of M\,31 and M\,33 ($783 \pm 25$\,kpc for M\,31 and $809 \pm 24$\,kpc for M\,33; \citealt{Mc2005}), it is challenging for the present ground and space telescopes to give spectroscopic observations for the individual M\,31 and M\,33 objects,
except for some very bright objects, such as the globular clusters and emission-line sources including
the planetary nebulae (PNe) and H~{\textsc{ii}} regions. 
The emission-line objects, whose energies are concentrated on a few emission lines,
 are excellent tracers for the study of the  kinematics and chemistry properties of their host galaxies. 

Many efforts have been done to identify and study the emission-line objects in M\,31 and M\,33. 
More than 30 years ago, Nolthenius \& Ford \citeyearpar{No1984} have identified 34 PNe in M\,31.
 \citet{Me2006} have presented a catalogue of 3,300 emission-line objects in the M\,31 area from the Planetary Nebular Spectrograph Survey.
2,730 of them are likely PNe, including 2,615 objects belonging to M\,31 and the remaining objects associating with the satellite galaxies and substructures of M\,31.  
\citet{Az2011} have presented a catalogue of 3,691 H~{\textsc{ii}} regions in M\,31 from the observations by the Mosaic Camera of the Mayall telescope and investigated the H$\alpha$ luminosity function of M\,31.
 \citet{Sa2012} have studied the metallicity profile of M\,31 based on the  spectra of hundreds of H~{\textsc{ii}} regions and PNe observed by the Multiple Mirror Telescope (MMT).
\citet{Lin2017} have observed 413 H~{\textsc{ii}}  regions in M\,33 with MMT and analyzed the detail temperatures and oxygen abundances distributions of M\,33. 
 \citet{Ma2018} have presented a catalogue of about 800 emission-line point-like sources in the central region of M\,31 based on the observation from Canada-France-Hawaii Telescope. 
\citet{Li2018} obtained optical spectra of 77 PNe in the circumnuclear region of M\,31 using the WIYN/Hydra multi-fiber spectrograph.

With 4000 fibers in a field of view of 20\,deg$^2$, the Large Sky Area Multi-Object Fiber Spectroscopic Telescope (LAMOST; \citealt{Cui2004}) 
is a powerful facility to archive spectra of millions of detectable objects. 
As an extension part of the LAMOST Spectroscopic Survey of the Galactic Anti-center (LSS-GAC; \citealt{Liu2014}; \citealt{Yuan2015}),  
the LAMOST M\,31/M\,33 survey targets interesting objects in the vicinity fields of M\,31 and M\,33, including the detectable objects in M\,31 and M\,33, such as the supergiants, PNe (\citealt{Yuan2010, Xiang2017}) , H~{\textsc{ii}} regions and globular clusters (\citealt{Chen2015, Chen2016}), 
background quasars (\citealt{Huo2010, Huo2013, Huo2015}) and foreground Galactic stars.

In this paper, we will present a systematic search of the H$\alpha$ emission-line point sources in the vicinity fields of M\,31 and M\,33,
based on the LAMOST M\,31/M\,33 survey data observed during October 2011 and January 2016.

The paper is organized as follows. In Section 2, we briefly introduce the LAMOST observation and data reduction.
Section 3 shows the selection of the H$\alpha$ emission-line point sources and Section 4 gives the results. Finally we summarize in Section 5.

\section{LAMOST OBSERVATION AND DATA REDUCTION}

Objects targeted by LAMOST in the M\,31 and M\,33 region include known and candidate objects in M\,31 and M\,33, 
such as PNe, supergiants, H~\textsc{ii} regions and globular clusters, known and candidate background quasars, and 
foreground Galactic stars. An overview of the target selection of the LAMOST M\,31/M\,33 survey 
can be referred to \citet{Yuan2015}. 
The LAMOST M\,31/M\,33 survey are targeted by LAMOST telescope from September to January of the next year in each observational season. 
The typical total exposure time, depending on the weather conditions, are 600 -- 1200\,s, 1200 -- 1800\,s and 1800 -- 2400\,s 
for bright (B), medium (M) and faint (F) plates, respectively.  
For most plates, the seeing varies between 3 and 4\,arcsec, with a typical value of about 3.5\,arcsec \citep{Yuan2015}. 

This work is based on spectra observed by LAMOST during September 2011 and January 2016. 
In total, 1,114,164 spectra have been collected by the LAMOST M\,31/M\,33 survey.  
Raw data was reduced by the LAMOST two-dimensional (2D) pipeline \citep{Luo2015}, including procedures of bias subtraction, 
cosmic ray removal, 1D spectral extraction, flat field correction, wavelength calibration and sky subtraction.  
The LAMOST spectra are separately recorded in blue (3700 -- 5900 ${\rm \AA}$) and red (5700 -- 9000 ${\rm \AA}$) arms. 
The blue and red spectra are processed by the 2D pipeline independently and combined together after the flux calibration,
which is processed by an algorithm specifically designed for the LSS-GAC survey \citep{Xiang2015}.
The final combined 1D spectra are then adopted to search for the H$\alpha$ emission-line objects and to obtain their properties in this work.

\section{Selections of H$\alpha$ Emission-Line Point Sources}

\begin{figure*}
 \centering
\includegraphics[width=0.8\textwidth]{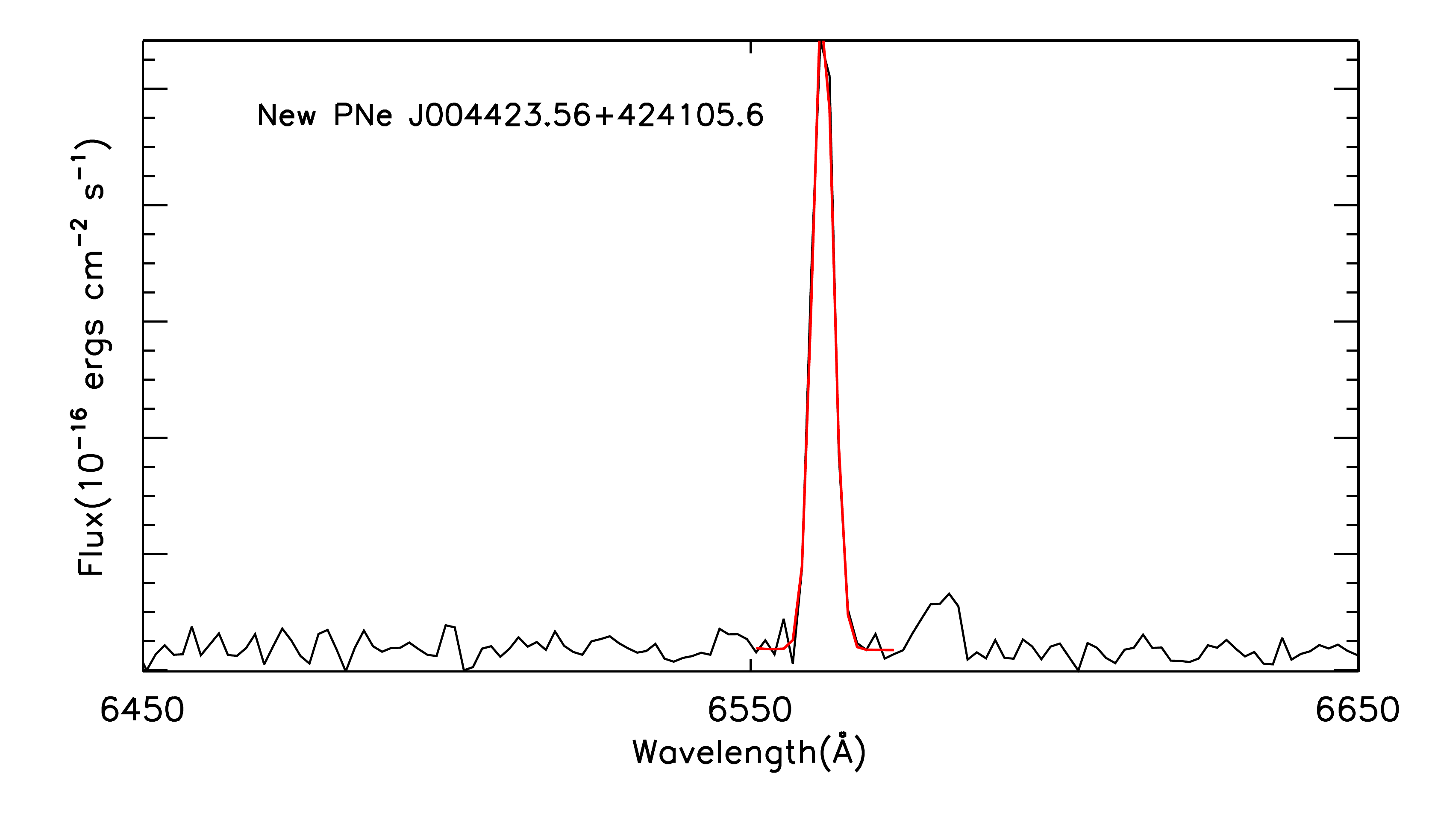}
   \caption{The Gaussian fitting of the H$\alpha$ emission line for object J00423.56+424105.6. 
   The black and red lines  show the LAMOST spectrum and the fitting results, respectively. }
\label{Fig1}
\end{figure*}

We first select objects with spectra that have H$\alpha$ emission-line by
the criteria of signal-to-noise ratio in the wavelength of 6563 $\rm \AA$ larger than 5. 
This yields a sample of 23,976 H$\alpha$ emission-line candidates. 
Their H$\alpha$ line profiles are then fitted by Gaussian functions and checked by naked eyes.
Any bad spectrum that has poor flat-fielding or sky subtraction is then excluded. 
As a result, this yields 4,448 unique objects as H$\alpha$ line emission-line sources.
Fig.~\ref{Fig1} shows an example of the Gaussian fitting of the H$\alpha$ emission line for an example object.

In the current work, we are only interested in the H$\alpha$ emission-line point sources. 
To exclude the extended sources such as the background galaxies and dwarf galaxies from our sample, 
we adopt the images and photometric catalogue of the Panoramic Survey Telescope and Rapid Response System-1 (Pan-STARRS~1; \citealt{Ka2002, Ka2010}).
795 candidates that are flagged as extended sources in the Pan-STARRS~1 catalogue are first excluded in our sample. 
 The Pan-STARRS~1 $r$-band images of all the remaining objects are visually examined to make sure that all sources in our sample are
 point sources and free from any contaminators.
 
There are usually two to three singular explosions for each LAMOST observation. Finally, we have visually examined
 the singular exposure spectra of the remaining sources and remove objects with fake H$\alpha$ emission line which 
appear in only one of those singular explosion spectra. As a result, we obtain 3,305 unique H$\alpha$ emission-line point sources as our final sample.

\section{Results}

\begin{table}
   \centering
  \caption{Description of the Catalogue.}
  \begin{tabular}{ccl}
  \hline
  \hline
Column & Name & Description  \\
\hline
1 & ID &  Object ID \\
2 & name & Object name \\
3 & RA & Right Ascension (J2000)  \\
4 & Dec  & Declination (J2000) \\
5  & plate &  LAMOST Plate ID \\
6 &  spid &  LAMOST Spectrograph ID   \\
7  & fbid &  LAMOST Fiber ID  \\
8 & mjd &  Observed MJD  \\
9 & objtype & Object type tagged in the input catalogue \\
10 & Vha & Radial velocity estimated from H$\alpha$ line \\
11 & Vhaerr & Uncertainty of radial velocity estimated from H$\alpha$ line \\
12 & fluxha & Flux of H$\alpha$ line \\
13 & fluxhaerr & Uncertainty of flux of H$\alpha$ line \\
14 & Vhb & Radial velocity estimated from H$\beta$ line \\
15 & Vhberr & Uncertainty of radial velocity estimated from H$\beta$ line \\
16 & fluxhb & Flux of H$\beta$ line \\
17 & fluxhberr & Uncertainty of flux of H$\beta$ line \\
18 & V4959 & Radial velocity estimated from [O~\textsc{iii}]~$\lambda4959$ line \\
19 & V4959err & Uncertainty of radial velocity estimated from [O~\textsc{iii}]~$\lambda4959$ line \\
20 & flux4959 & Flux of [O~\textsc{iii}]~$\lambda4959$ line \\
21 & flux4959err & Uncertainty of flux of [O~\textsc{iii}]~$\lambda4959$ line \\
22 & V5007 & Radial velocity estimated from [O~\textsc{iii}]~$\lambda5007$ line \\
23 & V5007err & Uncertainty of radial velocity estimated from[O~\textsc{iii}]~$\lambda5007$ line \\
24 & flux5007 & Flux of [O~\textsc{iii}]~$\lambda5007$ line \\
25 & flux5007err & Uncertainty of flux of [O~\textsc{iii}]~$\lambda5007$ line \\
26 & V6548 & Radial velocity estimated from [N~\textsc{ii}]~$\lambda6548$ line \\
27 & V6548err & Uncertainty of radial velocity estimated from [N~\textsc{ii}]~$\lambda6548$ line \\
28 & flux6548 & Flux of [N~\textsc{ii}]~$\lambda6548$ line \\
29 & flux6548err & Uncertainty of flux of [N~\textsc{ii}]~$\lambda6548$ line \\
30 & V6583 & Radial velocity estimated from [N~\textsc{ii}]~$\lambda6583$ line \\
31 & V6583err & Uncertainty of radial velocity estimated from [N~\textsc{ii}]~$\lambda6583$ line \\
32 & flux6583 & Flux of [N~\textsc{ii}]~$\lambda6583$ line \\
33 & flux6583err & Uncertainty of flux of [N~\textsc{ii}]~$\lambda6583$ line \\
34 & V6717 & Radial velocity estimated from [S~\textsc{ii}]~$\lambda6717$ line \\
35 & V6717err & Uncertainty of radial velocity estimated from [S~\textsc{ii}]~$\lambda6717$ line \\
36 & flux6717 & Flux of [S~\textsc{ii}]~$\lambda6717$ line \\
37 & flux6717err & Uncertainty of flux of [S~\textsc{ii}]~$\lambda6717$ line \\
38 & V6731 & Radial velocity estimated from [S~\textsc{ii}]~$\lambda6731$ line \\
39 & V6731err & Uncertainty of radial velocity estimated from [S~\textsc{ii}]~$\lambda6731$ line \\
40 & flux6731 & Flux of [S~\textsc{ii}]~$\lambda6731$ line \\
41 & flux6731err & Uncertainty of flux of [S~\textsc{ii}]~$\lambda6731$ line \\
42 & type &  Object classification of this work \\
\hline
\end{tabular} 
\end{table}

\begin{table}
   \centering
  \caption{Summary of the Catalogue}
  \begin{tabular}{llc}
  \hline
  \hline
& Type & Number \\
\hline
Emission stars &  New symbiotic star candidate & 1 \\
                        &  Others (M dwarfs and B[e] stars, etc.) & 1486 \\
\hline
Emission Nebulae &  Literature PNe & 353 \\
                              & New PN candidates & 24 \\
                              & Literature H~{\sc ii} regions & 64 \\  
                              & New  H~{\sc ii} region candidates & 19 \\  
                              & Literature SNRs & 10 \\  
                              & New  SNR candidates & 10 \\        
                              & Objects with  [O~{\sc iii}]~$\lambda$5007~flux $< $~3 & 52 \\                       
\hline
Others & Galaxies and unknown objects & 1286 \\
\hline
\end{tabular} 
\end{table}

\begin{figure*}
   \centering
   \includegraphics[width=0.8\textwidth]{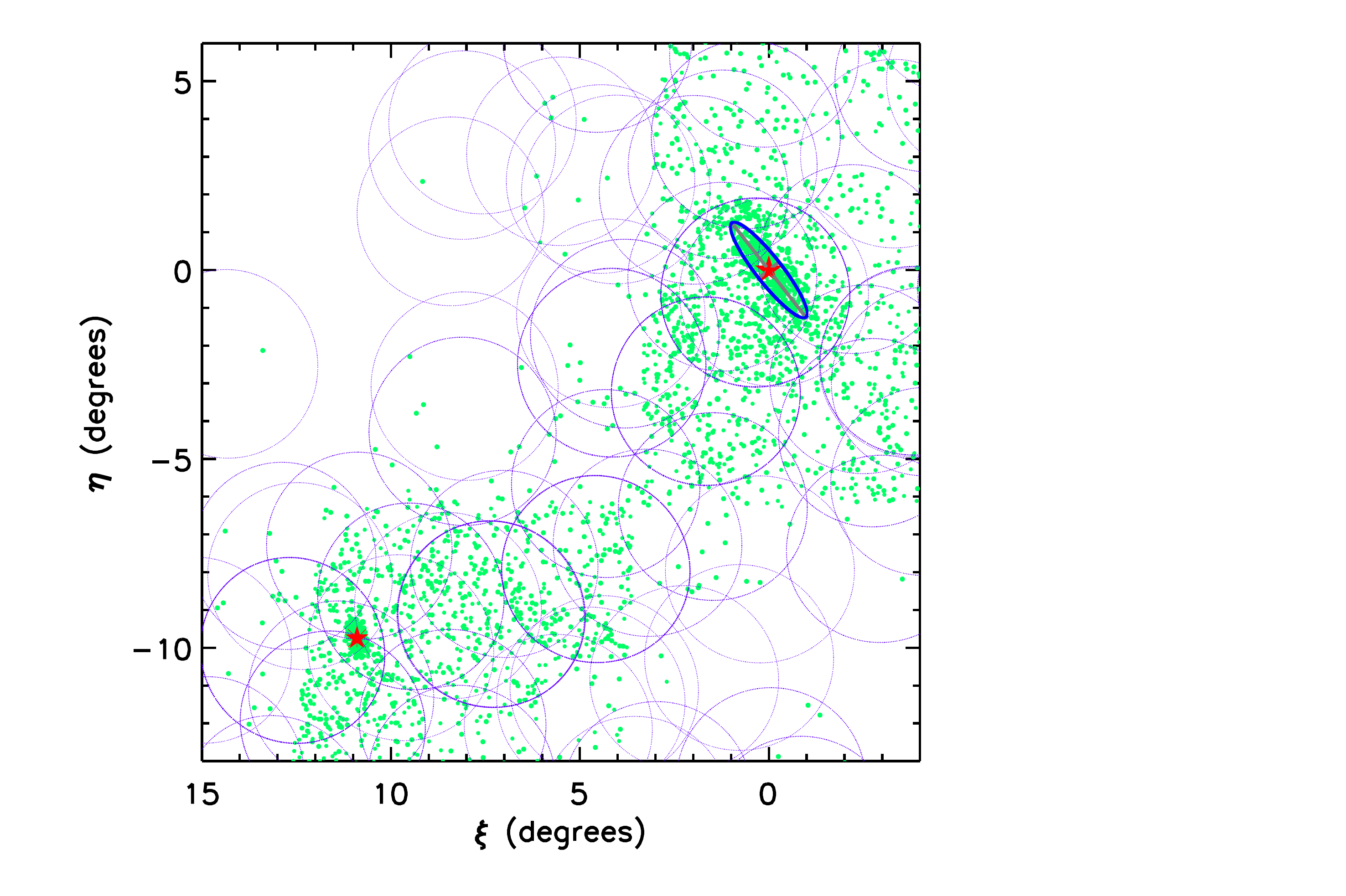}
   \caption{Spatial distributions of H$\alpha$ emission-line point sources in the vicinity fields of M\,31 and M\,33. 
   The purple circles mark the LAMOST plates. The red stars show the center of M\,31 and M\,33, respectively. The blue elliptic with a semi-major radius of 95.3 arcmin marks the disk region of M\,31.}
   \label{Fig3}
   \end{figure*}
   
\begin{figure*}
 \centering
  \includegraphics[width=0.8\textwidth]{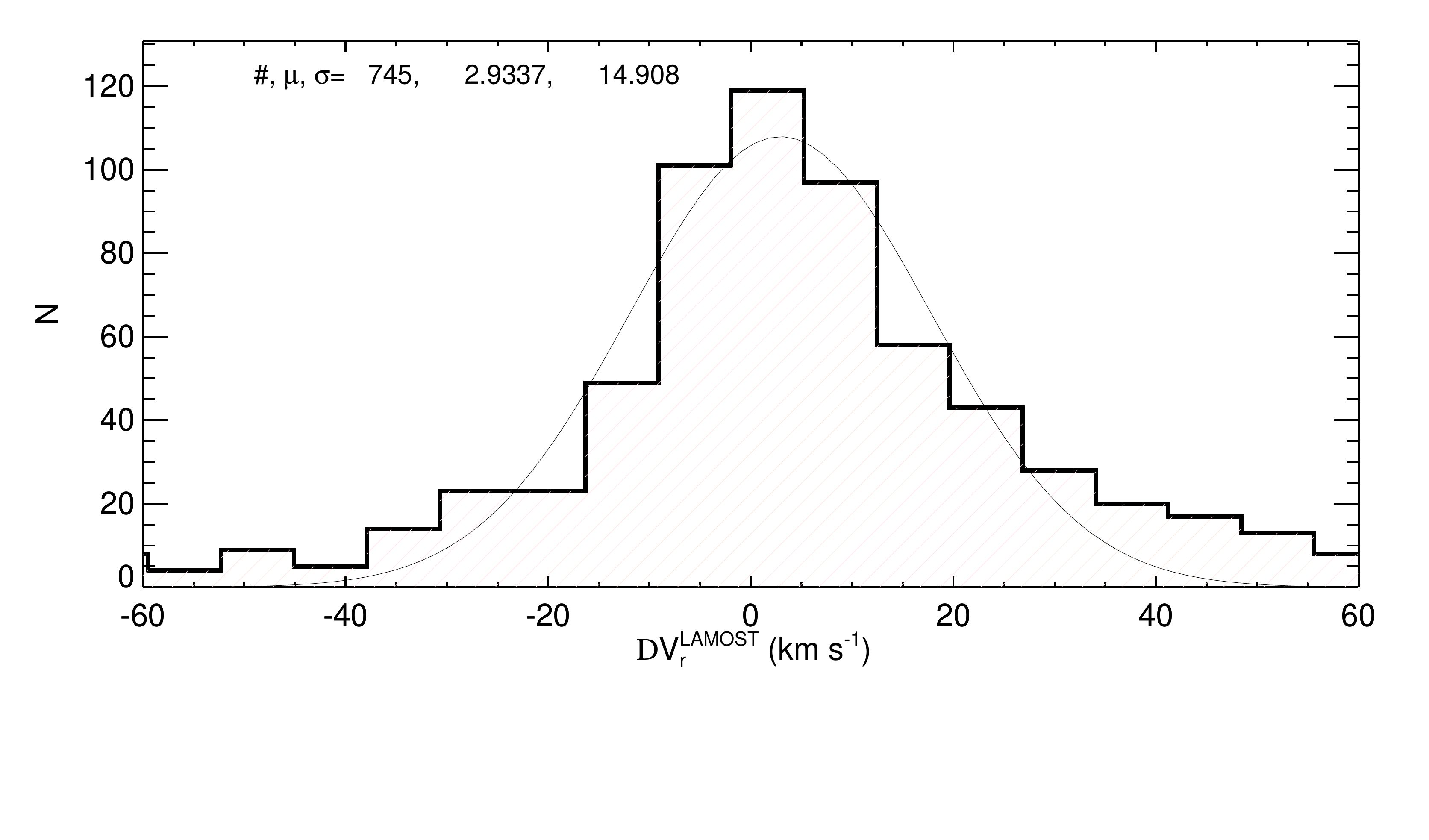}
 \caption{Histogram of the differences of radial velocities from H$\alpha$ line deduced from duplicate observations of the same targets. A Gaussian fit to the distribution is overplotted, with the number of objects with duplicate observations, and the mean and dispersion of the Gaussian marked.}
 \label{Fig4}
\end{figure*}

Our catalogue is available in electronic form in the online version of this  manuscript. Table~1 describes
the data format of the catalogue. 
Each row of the catalogue contains the ID, name, coordinates and LAMOST observational information, such as the Plate ID, Spectrograph ID, Fiber ID, 
observed MJD and the object type tagged in the input catalogue, of the catalogued objects.
Multiple observations of the same object are listed together with the same ID and object name.
Fig.\,\ref{Fig3} shows the spatial distribution of all the catalogued H$\alpha$ emission-line point sources in 
the vicinity fields of M\,31 and M\,33 in the $\xi$--$\eta$ plane. 
Here $\xi$ and $\eta$ are respectively the right ascension and declination offsets relative to the optical center of M\,31.

For all the catalogued objects, we have made Gaussian fits to their H$\alpha$ emission lines.
If there are other prominent emission lines such as  H$\beta$, [O~{\textsc{iii}}]~$\lambda$4959, 
[O~{\textsc{iii}}]~$\lambda$5007,  [N~{\textsc{ii}}]~$\lambda$6548,  [N~{\textsc{ii}}]~$\lambda$6583, 
[S~{\textsc{ii}}]~$\lambda$6717 and [S~{\textsc{ii}}]~$\lambda$6731
lines in the spectra, the Gaussian profile fits are also applied to these lines. 
Some of the emission lines (e.g. H$\alpha$ and [N~{\textsc{ii}}]~$\lambda$6583 lines) locate in 
narrow wavelength ranges. In those cases, the spectra are fitted by multiple Gaussian profiles, with one Gaussian component corresponding to one emission line.
The center wavelength and integrated area of the Gaussian profiles define the radial velocities and fluxes of the corresponding lines. 
The resulted properties such as the radial velocity and flux of the H$\alpha$ and seven other prominent emission lines 
(H$\beta$, [O~{\textsc{iii}}]~$\lambda$4959, [O~{\textsc{iii}}]~$\lambda$5007,  [N~{\textsc{ii}}]~$\lambda$6548,  [N~{\textsc{ii}}]~$\lambda$6583, 
[S~{\textsc{ii}}]~$\lambda$6717 and [S~{\textsc{ii}}]~$\lambda$6731) are also listed in our catalogue. 

There are 745 duplicate observations for our catalogued objects.
In Fig\,\ref{Fig4} we show the distribution of the differences of radial velocities from the H$\alpha$ line deduced from the duplicate observations. 
The rms of the differences is about 15 km\,s$^{-1}$, 
which indicate a internal uncertainty of about 10 km\,s$^{-1}$ for the H$\alpha$ line radial velocities.

Our H$\alpha$ emission-line point source catalogue consists of objects of different types, including 
the emission-line object such as  PNe, H~{\textsc{ii}} regions and supernovae remnants (SNRs) in M\,31 and M\,33 
and the Galactic emission-line stars. We cross-match our sample with the
SIMBAD Astronomical Database and find 575 of them been classified,
including 213 PNe, 161 emission-line stars and some other types of objects such as H~{\textsc{ii}} regions and SNRs.
However, as some of the SIMBAD sources are identified by only photometric data.
Their classification would be incorrect. In the current work, we will provide new classifications to the catalogued objects 
based on information of the LAMOST spectroscopic observations.
The new classifications are also listed in the catalogue and a summary is given in Table~2.

\subsection{Emission-Line Stars}

\begin{figure*}
   \centering
   \includegraphics[width=0.9\textwidth]{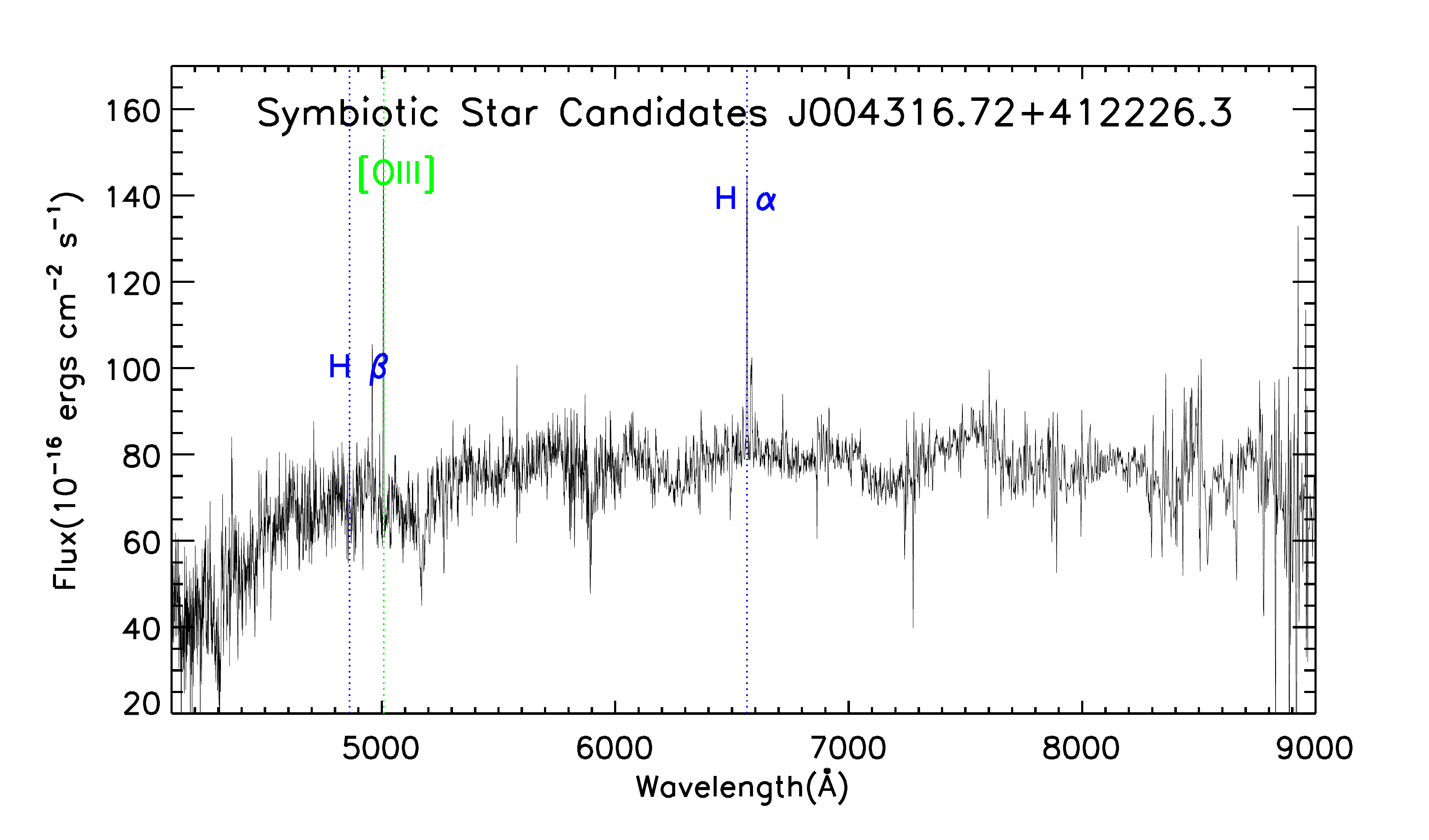}
   \caption{LAMOST spectrum of a possible symbiotic star LAMOST J004316.72+412226.3.
   Vertical lines with different colors mark the positions of the different emission lines.}
   \label{Fig5}
   \end{figure*}

Based on their LAMOST spectra, the catalogued objects can be roughly divided into two groups, 
one group of objects with significant continuum spectrum and the other group without.
There are 1,487 objects in our catalogue belonging to the first group. 
Most of them have spectra of typical stars and have only the H$\alpha$ emission lines.
We classify them as emission-line stars in the current work.
Most of the emission-line stars are M dwarfs displaying strong TiO bands. 
There are also several B[e] stars, which are B-type main-sequence sub-giant or giant stars with prominent Balmer emission, and some novae.

One of our catalogued object, LAMOST J004316.72+412226.3, has a spectrum that is very likely to be a symbiotic star. 
Symbiotic stars are interacting binaries which usually contain a white dwarf and a red giant \citep{Mi2014}.
The spectrum of a symbiotic star usually  presents both the feature of a late-type M giant and strong emission lines 
such as the Balmer H~{\textsc{i}} lines and the [O~{\textsc{iii}}]~$\lambda$5007 forbidden lines.
The possible symbiotic star LAMOST J004316.72+412226.3 we find in the current work was classified as a PN in \citet{Ha2006}.
Its LAMOST spectrum is displayed in Fig.\,\ref{Fig5}.
The spectrum shows typical features of M stars in the red band and hot stars in the blue band. 
The emission lines of H$\alpha$  and  [O~{\textsc{iii}}]~$\lambda$5007 lines are significant.
However, due to the low resolution and limited signal-to-noise ratio of the LAMOST spectrum, follow up observations of this object are 
needed for any further confirmation.
   
\subsection{Emission-Line Nebulae}

 \begin{figure*}
 \centering
   \includegraphics[width=0.45\textwidth]{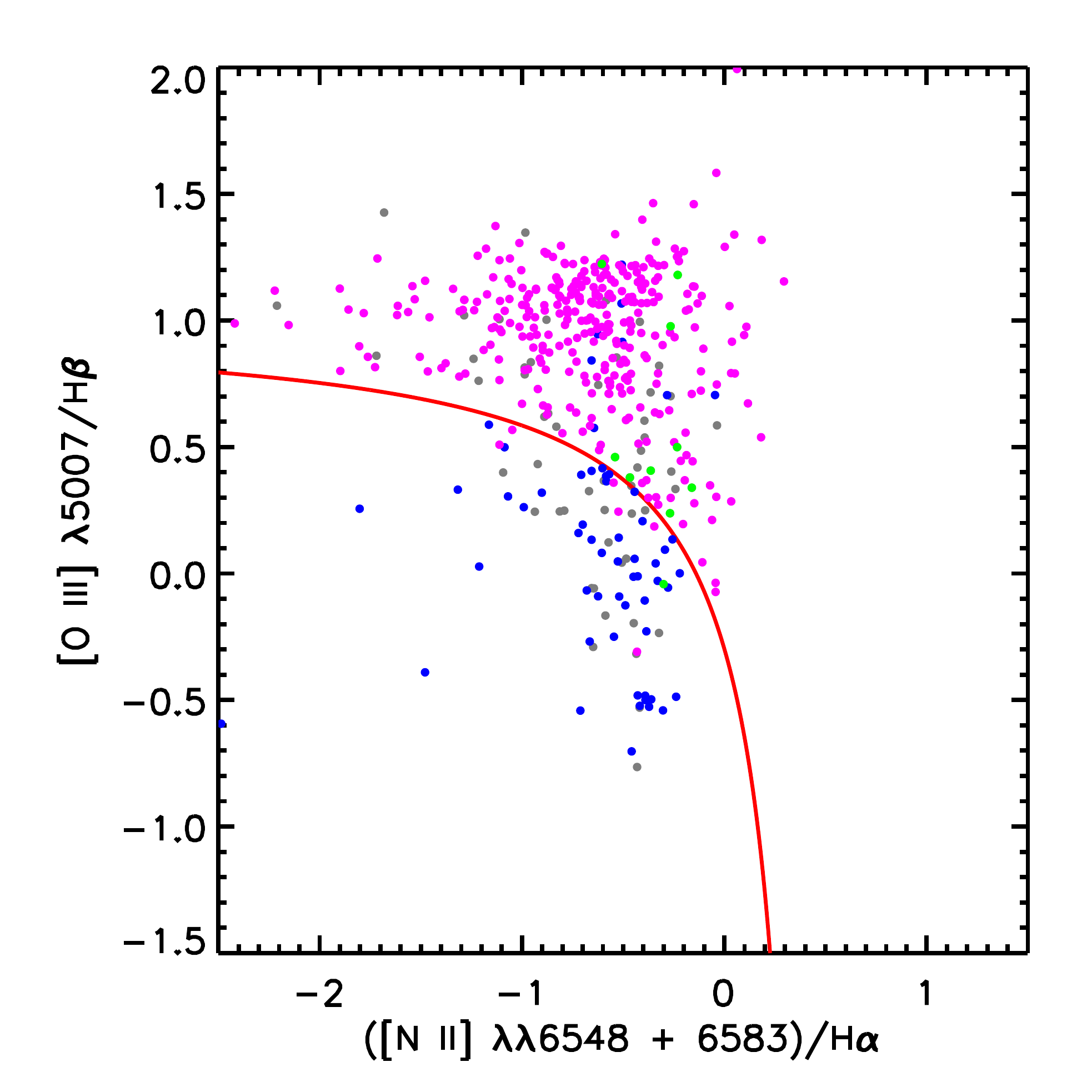}
   \includegraphics[width=0.45\textwidth]{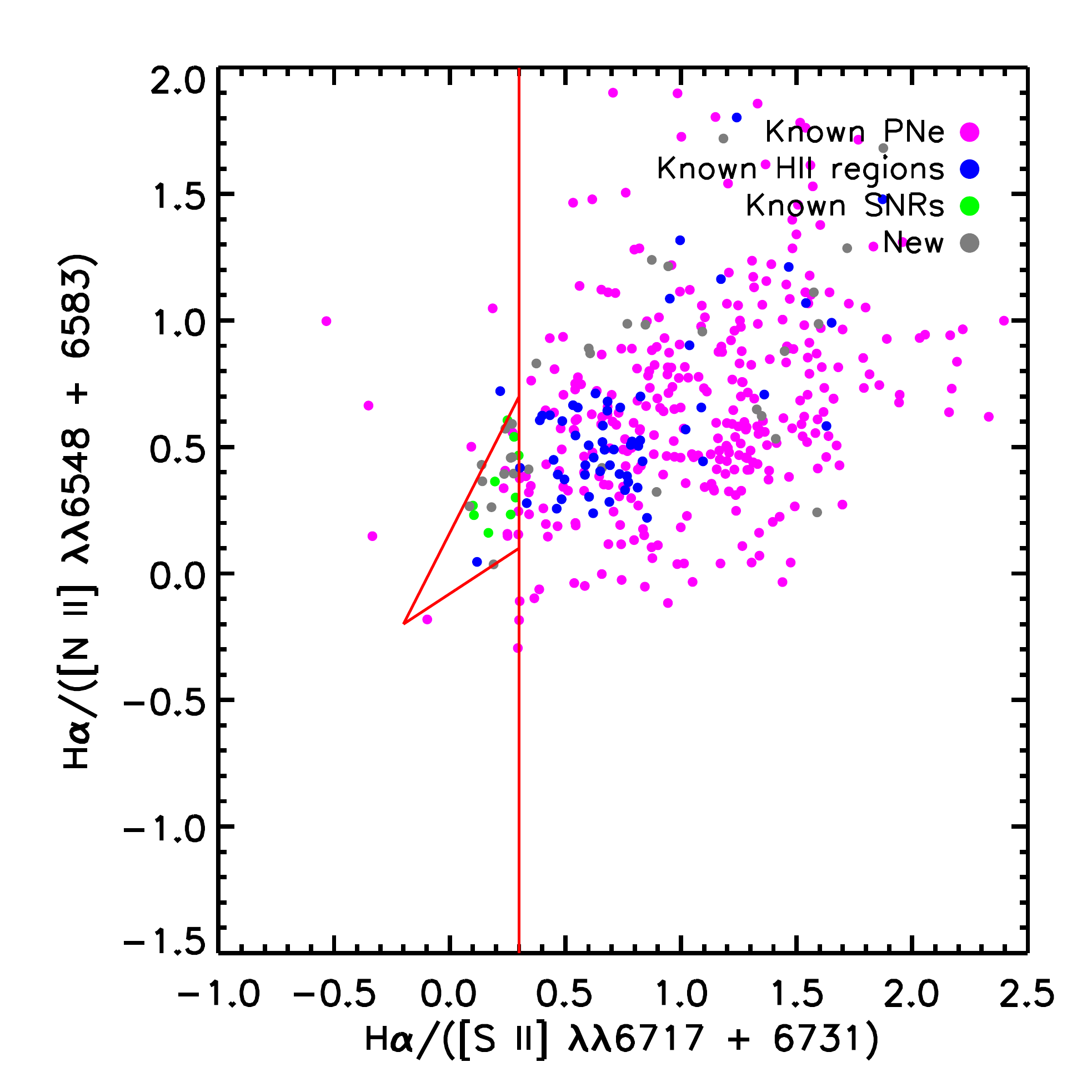}
  \caption{The BPT (left) and SMB (right) diagrams of our catalogued emission-line nebulae with strong [O~{\textsc{iii}}]~$\lambda$5007 emission (grey dots). 
 Known PNe, H~{\textsc{ii}} regions and SNRs are plotted as pink, blue and green dots, respectively. 
The red line in the left panel marks the line we adopted to separate PNe and H~{\textsc{ii}} regions while
those in the right panel are the criterions we adopted to select the SNR candidates  (see text for details).}
  \label{Fig6}
\end{figure*}

\begin{figure*}
   \centering
   \includegraphics[width=0.8\textwidth]{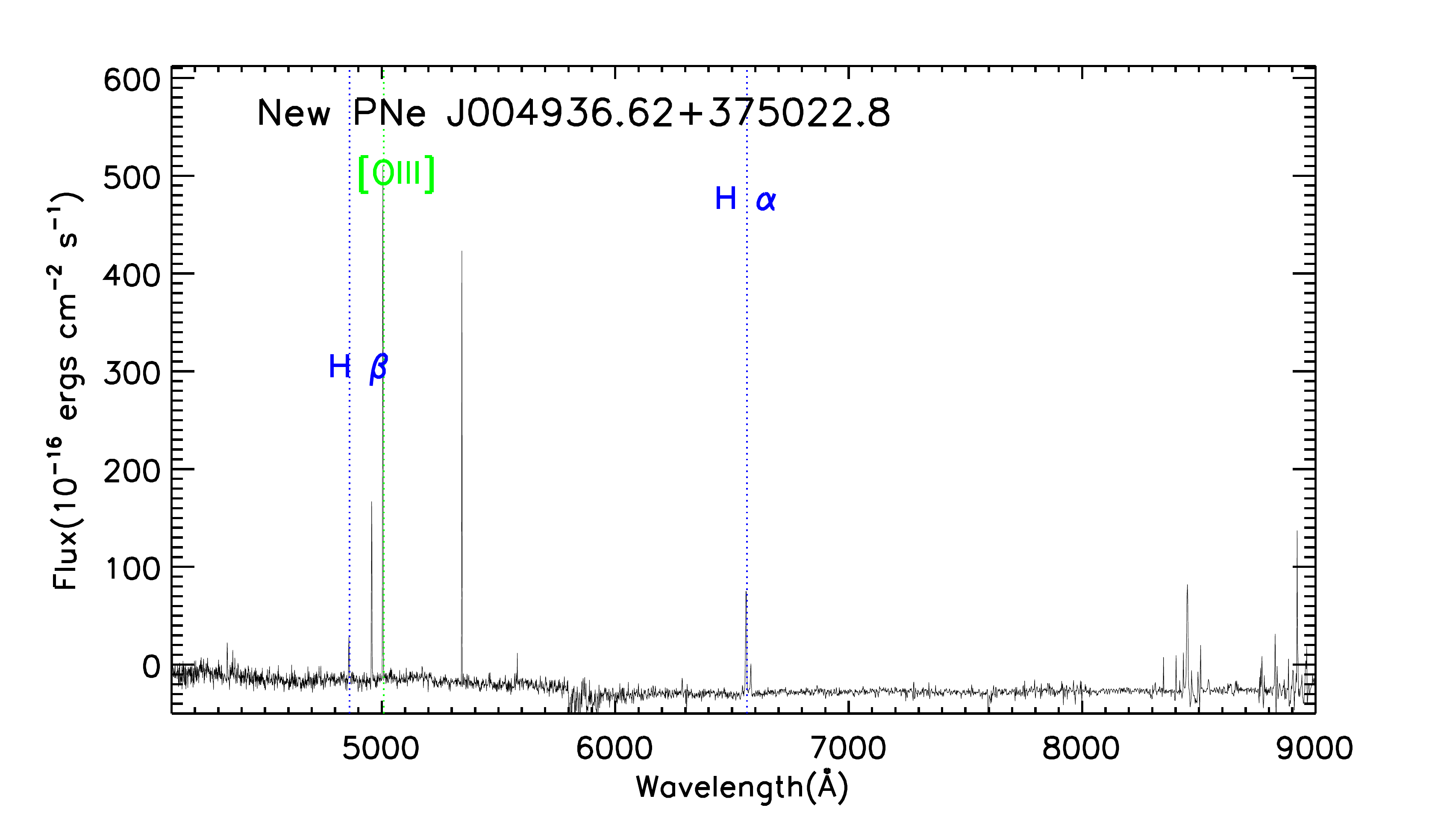}
    \includegraphics[width=0.8\textwidth]{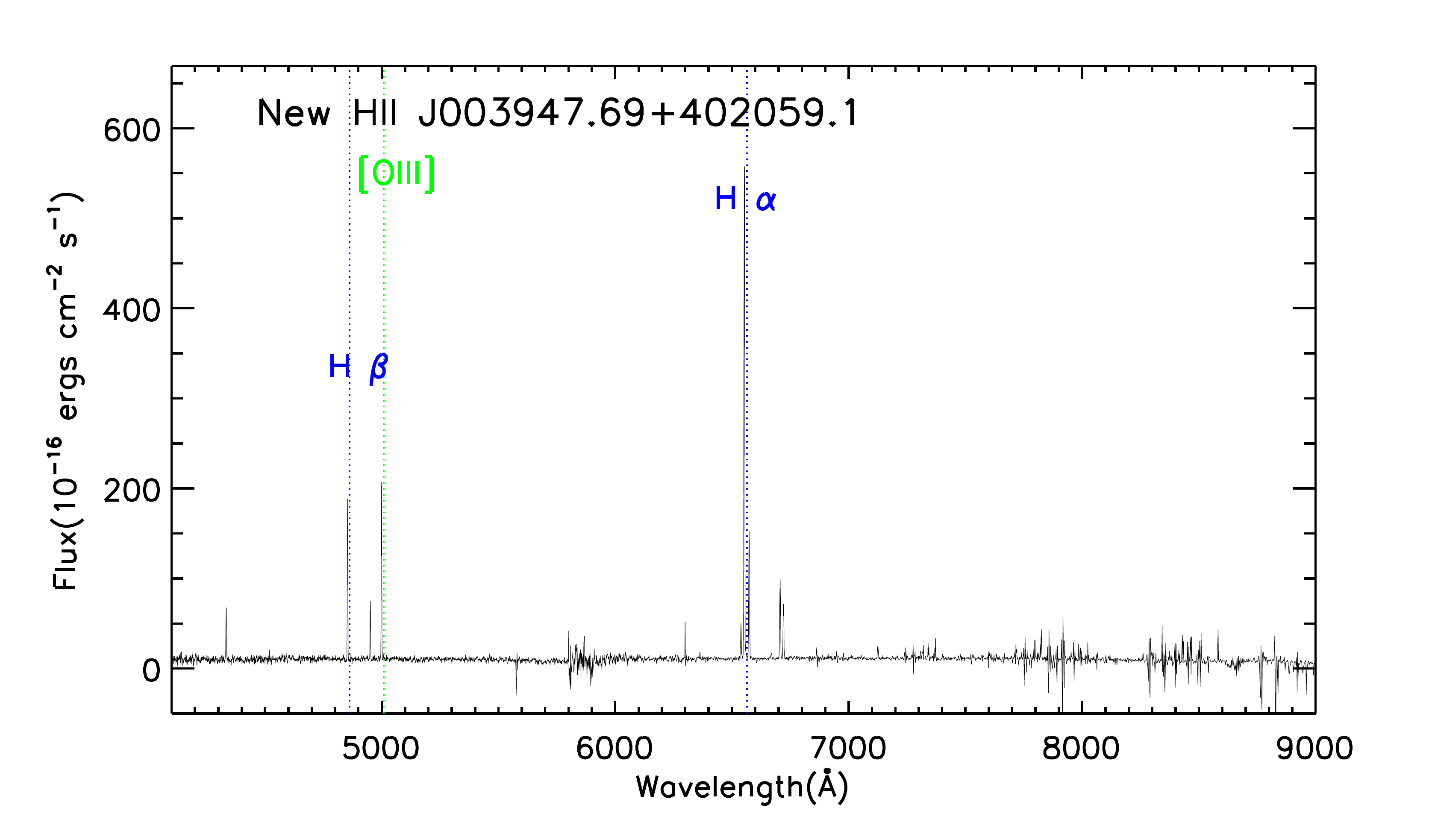}
   \caption{ LAMOST spectra of a PN candidate LAMOST J004936.62+375022.8 (upper panel) and a
   H~{\textsc{ii}} region candidate LAMOST J003947.69+402059.1 (bottom panel). 
   Vertical lines with different colors mark the positions of the different emission lines.}
   \label{Fig7}
   \end{figure*}
   
   \begin{figure*}
   \centering
   \includegraphics[width=0.69\textwidth]{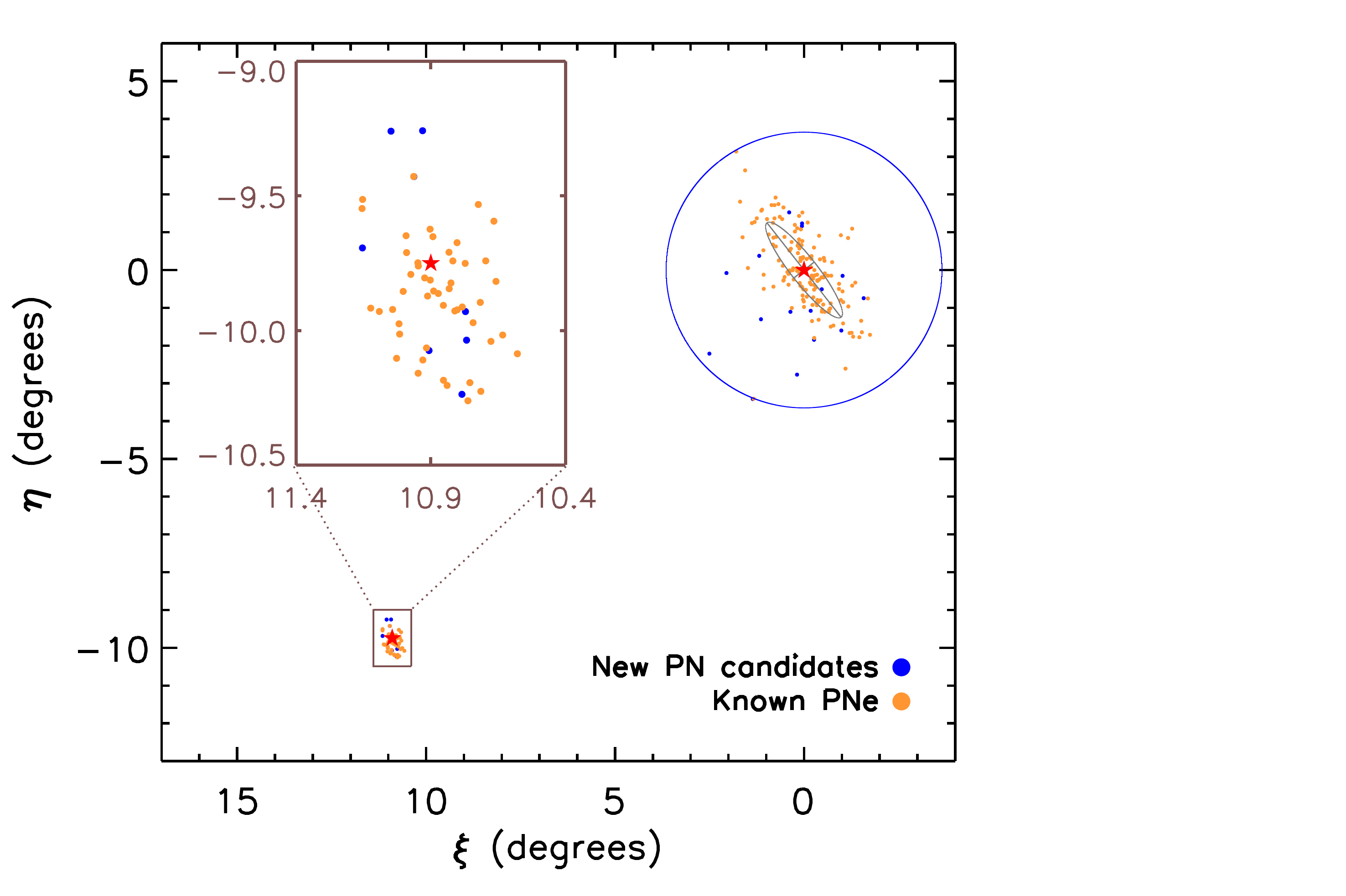}
   \caption{The spatial distribution of the new PN candidates (blue dots) and known PNe from SIMBAD (orange dots) in our catalogue. 
   Red star symbols mark the center of the M\,31 and M\,33 and the grey ellipse the M\,31 disk region with a semi-major radius of a semi-major radius of 95.3\,arcmin.
   The blue circle shows the region with a projected distance of 50\,kpc to the M\,31 center.}
   \label{Fig8} 
   \end{figure*}

\begin{figure*}
   \centering
    \includegraphics[width=0.69\textwidth]{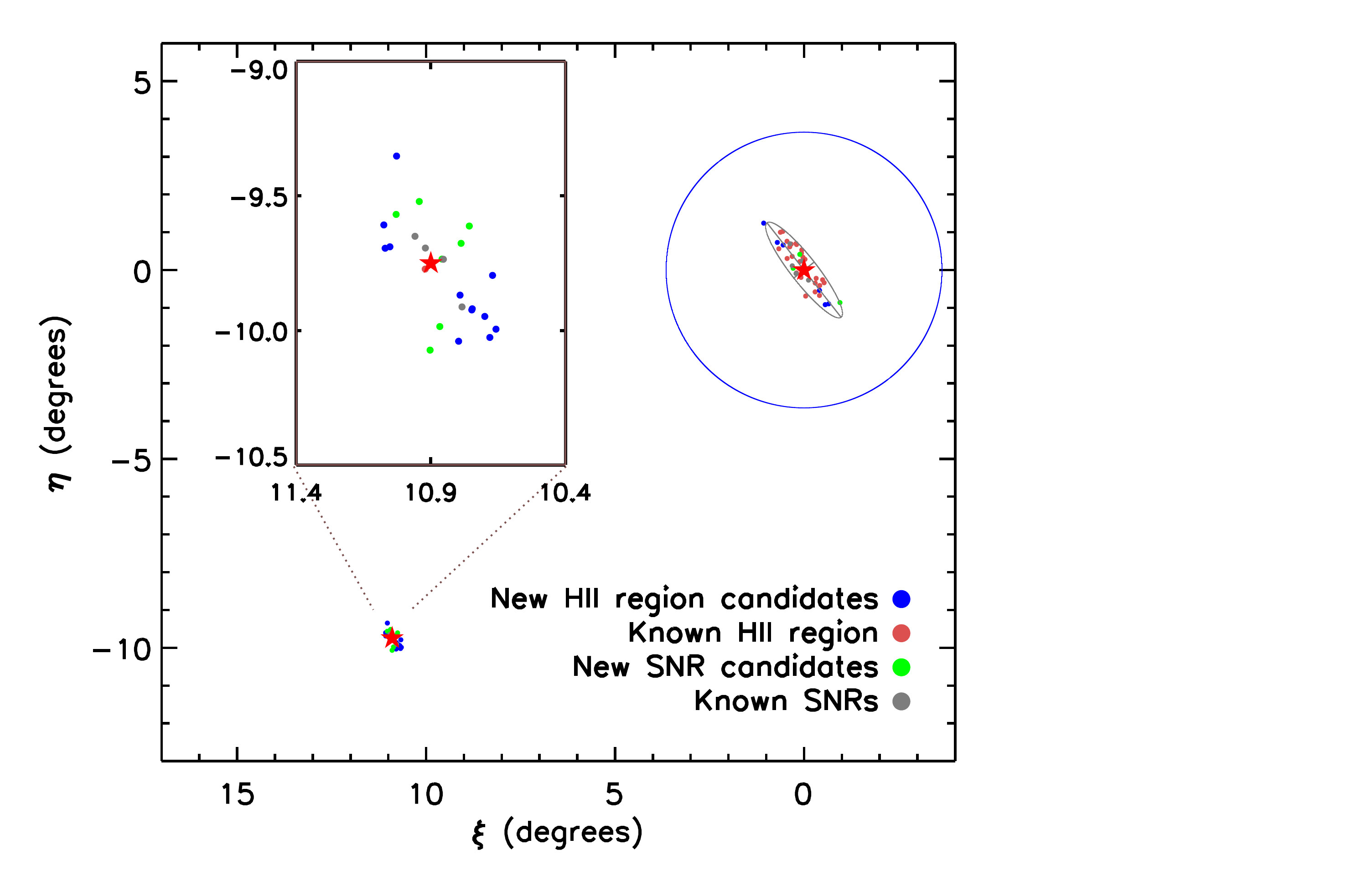}
   \caption{Similar as Fig.\,\ref{Fig8} but for H~{\textsc{ii}} regions and SNRs.  
   Known and newly identified candidate  H~{\textsc{ii}} regions, as well as  known and 
   new SNRs candidates are plotted with brown, blue grey and green dots, respectively. }
   \label{Fig11}
   \end{figure*}

In our catalogue, there are 532 objects which display no obvious continuum spectra but have abundant emission lines.
They are possible PNe, H~{\textsc{ii}} regions, and SNRs. 

To exclude the contamination of early type stars, in the current work, we classify only the sources with strong emission of [O~{\textsc{iii}}]~$\lambda$5007 line.
These are objects of high electron excitation states, i.e. PNe, H~{\textsc{ii}} regions and SNRs.
In the current work, we have selected 480 sources with [O~{\textsc{iii}}]~$\lambda$5007 fluxes greater than 3. 
To distinguish the PNe, H~\textsc{ii} regions and SNRs from each others, we adopt the commonly used ``Baldwin, Phillips \& Terlevich'' (BPT) and 
``Sabbadin, Minello \& Biancini'' (SMB) diagrams.
The BPT diagram, which demonstrate how PNe can be distinguished from H~{\textsc{ii}} regions on the basis of
 their [O~{\textsc{iii}}]~$\lambda$5007/H$\beta$ (noted as `O3') and  
 ([N~{\textsc{ii}}]~$\lambda$$\lambda$6548 + 6583)/H$\alpha$ (noted as `N2') flux ratios (\citealt{Ba1981}), 
 is then adopted to classify the PNe and H~{\textsc{ii}} region candidates in our catalogue. 
Comparing to PN and H~{\textsc{ii}} region, the sulfur in SNR is usually found in a wide range of 
ionization states and the nitrogen emission line of SNR is sensitive to the presence of its radiative shock.
We thus adopt the SMB diagram which separate the SNRs from PNe and H~{\textsc{ii}} regions by H$\alpha$/([N~{\textsc{ii}}]~$\lambda$$\lambda$6548 + 6583)
(noted as `n2') and H$\alpha$/([S~{\textsc{ii}}]~$\lambda$$\lambda$6717 + 6731) (noted as `s2') flux ratios (\citealt{Sa1977}; \citealt{Ca1981} and \citealt{Ri2006}).
Fig.\,\ref{Fig6} shows the BPT and SMB diagrams  of our catalogued emission-line nebulae with strong emission lines of [O~{\textsc{iii}}]~$\lambda$5007.
As a result, we have identified 377 likely PNe, 83 possible H~\textsc{ii} regions and 20 SNRs candidates in our catalogue.

 \subsubsection{Planetary Nebulae} 

The curve ${\rm O3} > 0.61/({\rm N2}-0.47) + 1.0$ \citep{Sa2012} separates the PNe and H~{\textsc{ii}} regions nicely (Fig.\,\ref{Fig6}).
It is adopted in the current work to select the PNe and we have identified 377 PN candidates in our catalogue, 
including 353 known PNe from the literature and 24 new discoveries.
The spectrum of an example PN candidate is shown in Fig.\,\ref{Fig7}. 
The spatial distribution of all the PN candidates is shown in Fig.\,\ref{Fig8}.
Most of the PN candidates are distributed in the disk regions of M\,31 and M\,33. 
However, some of the PN candidates, especially those newly discovered ones, locate in the halo regions of M\,31 and M\,33. 
They are valuable for us to study the chemistry and kinematics of the halos of M\,31 and M\,33 as well as streams in the halos.

\subsubsection{H~{\textsc{ii}} Regions}

From the BPT diagram, we have identified 83 H~{\textsc{ii}} region candidates in our catalogue, including 64 known objects from the literature 
and 19 new identifications.
The LAMOST spectrum of an example H~{\textsc{ii}} region candidate is also shown in Fig.\,\ref{Fig7} and
the spatial distribution of all the catalogued H~{\textsc{ii} region candidates is shown in Fig.\,\ref{Fig11}.

\subsubsection{Supernovae Remnants}

Following the work of Frew et al.\citeyearpar{Fr2010}, we adopt criteria: $0.6 \log {\rm s2} - 0.08 < \log {\rm n2} < 1.8 \log {\rm s2} +0.16$ and  
$\log {\rm n2} < 0.3$ to select the SNRs candidates (Fig.\,\ref{Fig6}). 
As a result, we have identified  20 SNRs candidates in the current work,
 including 10 known ones from literature and 10 new discoveries. Their spatial distribution is also shown in Fig.\,\ref{Fig11}.

\section{Summary}

Based on the spectroscopic data from LAMOST M\,31/M\,33 survey during September, 2011 and June, 2016, 
we have selected 3,305 H$\alpha$ emission-line point sources in the vicinity of M\,31 and M\,33.
We have calculated radial velocities and fluxes of H$\alpha$ line and those quantities of other 
7 major emission lines (H$\beta$, [O~{\textsc{iii}}]~$\lambda$4959, [O~{\textsc{iii}}]~$\lambda$5007,  [N~{\textsc{ii}}]~$\lambda$6548,  [N~{\textsc{ii}}]~$\lambda$6583, 
[S~{\textsc{ii}}]~$\lambda$6717 and [S~{\textsc{ii}}]~$\lambda$6731) of all these catalogued sources from their LAMOST spectra by Gaussian profile fitting method.
In our catalogue, we have identified 1487 emission-line stars, including 1 new symbiotic star candidate, 
532 emission-line nebulae, including 24 new PN candidates, 19 new H~{\textsc{ii}} regions candidates and 10 new SNRs candidates, and 1286 unknown objects. 
Our catalogue is available in electronic form in the online version of this  manuscript. 
It will serve as an ideal starting point to study the chemistry properties and kinematics of the M\,31.

\begin{acknowledgements}
This work was funded by the National Natural Science Foundation of China (NSFC)
under No.11080922, 11803029, 11973001, U1531244, 11833006 and U1731308 and 
National Key R\&D Program of China No. 2019YFA0405500.

This work has made use of data products from the Guoshoujing Telescope (the Large Sky Area Multi-Object Fibre Spectroscopic Telescope, LAMOST). 
LAMOST is a National Major Scientific Project built by the Chinese Academy of Sciences. Funding for the project has been provided by the National Development and Reform Commission. 
LAMOST is operated and managed by the National Astronomical Observatories, Chinese Academy of Sciences.
\end{acknowledgements}

\bibliographystyle{raa}
\bibliography{msRAA-2019-0363}

\end{document}